\documentstyle[twocolumn,aps]{revtex}
\begin{document}
\draft
\title{The interplay between electron-electron interactions
and impurities in one-dimensional rings}
\author{Hiroyuki Mori\cite{leave}}
\address{Department of Physics, Indiana University, Bloomington, IN 47405}
\date{\today}
\maketitle
\begin{abstract}
The persistent current and charge stiffness of a one-dimensional
Luttinger liquid on a ring threaded by a magnetic flux
are calculated by
Monte Carlo simulation. By changing the random impurity potential
strength and the electron-electron interaction, we see a crossover
behavior between weak and strong impurity limits.
For weak impurity potentials, interactions enhance impurity effects,
that is, interactions decrease the current and the stiffness.
On the other hand, interactions tend to screen impurities
when the impurity potential is strong.
Temperature dependence of the persistent current
and the charge stiffness shows a peak at a characteristic temperature,
consistent with a recent single impurity study.
\end{abstract}
\narrowtext
Recent technological progress now allows us to produce nearly ideal
low-dimensional systems, and it is becoming more important
to understand how disorder and electron-electron
interactions control physics in such systems.
In one-dimensional systems,
Giamarchi and Schultz\cite{giamarchi} derived
renormalization group (RG) equations for systems
with weak but finite density of impurities, and
showed that the critical impurity strength is zero
for the Luttinger liquid parameter $g>2/3$ and is finite
for $g<2/3$, where $g>1$ and $g<1$ correspond to repulsive and
attractive interactions, respectively.
Scalettar {\it et al.}\cite{scalettar} confirmed a part of
the phase diagram given by Giamarchi and Schultz, using quantum
Monte Carlo (MC) simulation of the boson Hubbard model.
\cite{doty}

For finite-size systems, current can flow even in
the presence of impurities for any $g$ because of the finite size.
One interesting example of a finite-size system
is a one-dimensional ring threaded by magnetic flux,
which creates persistent current as first predicted by
Byers and Yang and later by B\"{u}ttiker {\it et al.}
\cite{byers}.
When an electron gas moves in a dirty ring, we expect
the persistent current to be suppressed as the impurity potential
becomes stronger \cite{cheung}. What then will happen if electron-electron
interactions are turned on \cite{ambegaokar}?
Krive {\it et al.}\cite{krive} calculated
persistent current of Wigner crystal on a ring with a single impurity
in certain limits. From their expressions we can see the current
decreases as we increase particle-particle repulsive interactions.
This study indicates that
interactions effectively enhance the impurity strength, although
their system has only one impurity.
For a finite density of weak impurities, we can use RG equations
obtained by Giamarchi and Schultz, placing a cutoff
in the length scale since the system is finite.
Then the effective impurity strength $\tilde{W}$ is given by $\tilde{W}\sim
L^{3-2/g}$, where $L$ is the system size and
$g$ is the Luttinger liquid parameter which
becomes large as repulsive interactions increase.
Here we can see again that interactions would enhance the effective
impurity strength.

There have been several exact diagonalization studies done
for the ring problem \cite{bouzerar}.
These calculations were performed in diffusive regime, {\it i.e.},
for relatively weak impurities, and
showed qualitatively the same results as the above analytical works.

This interplay between interactions and impurities can be understood
in the following way. When impurities are weak enough, the system still
resides in the Luttinger liquid regime \cite{this}. Luttinger liquids
have charge-density wave (CDW) correlations which grow
as the repulsive interaction becomes strong. Since a CDW would be easily
pinned by impurities, stronger interactions make the system
pinned more easily. Therefore the persistent current
would decrease when interactions increase.

Now the question is what will happen with strong impurities.
It seems clear that
interactions would {\it screen} the impurities,
which is opposite to what is expected in the weak impurity limit.
When the impurities are very strong,
particles are almost localized in the low potential regions.
If we turn on repulsive interactions between particles,
it will move the particles away from each other, and therefore
the interactions would effectively weaken the impurity strength.

In this letter we will show the crossover behavior
between these weak and strong impurity limits.
We have calculated the persistent current and charge stiffness
of a spinless Luttinger liquid on a ring with a finite density of impurities,
using Monte Carlo simulation. One advantage of working with a  Luttinger
liquid instead of a microscopic lattice model is that we can separate out
the lattice (periodic potential) effect, which is another important factor
changing those quantities \cite{muller}, so we can concentrate on the
interplay between impurities and interactions only \cite{there}.
We also show the temperature dependence of persistent current
and charge stiffness. These have peaks at certain temperatures,
as previously argued for the single impurity problem\cite{krive}.

The problem of Luttinger liquid on an impurity-free ring
penetrated by a magnetic flux was solved by Loss \cite{loss}.
He showed that the Lagrangian is given at $T\ll \pi g_{x} /2L$  by\\
\begin{equation}
{\cal L}=\frac{g_{\mu}}{8\pi}(\partial_{\mu}\theta)^{2}+
(N-1+2\phi/\phi_{0})\frac{i}{2L}\partial_{\tau}\theta,
\end{equation}
where $\mu=\tau$ and $x$.
The boundary condition of $\theta$ field is
$\theta(x+L,\tau +\beta)=\theta(x,\tau) + 2n\pi$,
where $n$ is a winding number.
The derivatives of $\theta$ give the particle density and current:
$\partial_{\mu}\theta = -2\pi j_{\mu}$.
Defining $\tilde{\theta}\equiv\theta - 2n\tau\pi/\beta,$ we obtain
the partition funtion $Z$ in the following form,
\begin{equation}
Z=\sum_{n}\int d\tilde{\theta}(-1)^{(N-1)n}\cos (2\pi n\frac{\phi}{\phi_0})
{\rm e}^{-\frac{T}{T_0}n^2}{\rm e}^{-\int \tilde{{\cal L}}}
\end{equation}
where $\tilde{{\cal L}}=(g_{\mu}/8\pi)(\partial_{\mu}\tilde{\theta})^{2},$
and $T_0=2/g_{\tau}L\pi,$ In the following calculation we assume the
particle number $N$ is odd, since we can get the physical
quantities for the cases where $N$ is even by using the relation
 $Z(N\ \mbox{even}; \phi/\phi_0)=Z(N\ \mbox{odd}; \phi/\phi_0+1/2)$.
Haldane \cite{haldane1}
calculated the relation of $g_{x}$ and $g_{\tau}$ to the
microscopic parameters of a spinless fermion system with nearest
neighbor interactions.
$g_{\tau}$ varies with density
but does not change very much with interaction
as long as the interaction is not too large, while $g_{x}$ changes
with both density and interaction.
The sound velocity $v_s$ is given by $\sqrt{g_{x}/g_{\tau}}$.
In the case of the noninteracting model, $g_x=1/g_{\tau}=v_s=v_F$.

The impurity term is written in the microscopic model as
$\int dxV_{\rm imp}(x)\rho(x)$.
Following Ref. \cite{haldane2}
we obtain
\begin{eqnarray}
\int V_{\rm imp}(x)\rho(x)&=&
\int\frac{1}{2\pi}V_{\rm imp}\partial_{x}\theta
+\int V_{\rm imp}(x)\rho_0{\rm e}^{i2k_{F}x}{\rm e}^{i\theta}\nonumber\\
&&+\mbox{h.c.}
+\mbox{higher orders},
\end{eqnarray}
where $\rho_{0}$ is average density.
The first term is forward scattering and the second represents
backward scattering.
Using $\tilde{\theta}$ instead of $\theta$ we obtain
\begin{eqnarray}
\tilde{{\cal L}}&=&\frac{g_{\mu}}{8\pi}(\partial_{\mu}\theta)^{2} +
\frac{1}{2\pi}V_{\rm imp}(x)\partial_{x}\theta\nonumber\\
&&+ 2\rho_0 V_{\rm imp}(x)\cos
(\theta+2\pi\rho_{0}x+2n\pi \tau /\beta),
\end{eqnarray}
where we have omitted the tilde from $\tilde{\theta}$ and
we used $k_{F}=\pi\rho_{0}$.
Since the forward scattering is irrelevant\cite{giamarchi},
we took into account only the first and third terms in our
calculations.

In order to do MC simulations, we do not want to have
the Aharonov-Bohm phase factor
$\cos (2\pi\phi/\phi_0)$ in Eq.(2) because it can give a negative
weight. Fortunately we can express persistent current for finite $\phi$
in terms of $\langle n^k\rangle|_{\phi =0}$ as will be shown below, so that
we just need to perform MC simulations at $\phi =0$ which causes no
negative weight problem.

Here we show how to calculate persistent current at finite $\phi$ using
quantities at $\phi =0$.
The persistent current $I$ on a ring is given by
$I(\phi)=-(1/\beta)\partial \ln Z(\phi)/ \partial\phi=%
-(2\pi i/\beta\phi_{0})\langle n\rangle$.
We expand $I(\phi)$ in a Fourier series:
$I(\phi)=\sum_{n}I_{n}\sin(2\pi n\phi/\phi_{0})$,
in other words,
$\langle n\rangle_{(\phi)}=%
\beta i(\phi_{0}/2\pi)\sum_{n}I_{n} \sin(2\pi n \phi/\phi_{0})$.
Taking derivatives of $\langle n\rangle$
with respect to $\phi$ and setting $\phi=0$,
we obtain,
\begin{equation}
\frac{\partial^{2k+1}\langle n\rangle}{\partial\phi^{2k+1}}|_{\phi=0}  =
(-)^{k}\beta i (\frac{\phi_{0}}{2\pi})^{2k}\sum_{n}I_n n^{2k+1},
\end{equation}
and $\partial^{2k}\langle n\rangle/\partial\phi^{2k}|_{\phi=0}=0$
On the other hand, from the definition of $\langle n\rangle$ we have
\begin{eqnarray}
\frac{\partial\langle n\rangle}{\partial\phi}|_{\phi=0} & =
& \frac{2\pi i}{\phi_{0}}\langle n^{2}\rangle|_{\phi=0},\\
\frac{\partial^{3}\langle n\rangle}{\partial\phi^{3}}|_{\phi=0} & =
& (\frac{2\pi i}{\phi_{0}})^{3}
(\langle n^{4}\rangle-3\langle n^{2}\rangle^{2})|_{\phi=0},
\end{eqnarray}
and $\partial^{2k}\langle n\rangle/\partial\phi^{2k}|_{\phi=0}=0$.
Consequently, we have
\begin{eqnarray}
\frac{2\pi}{\phi_{0}}\langle n^{2}\rangle|_{\phi=0} & = &
\beta\sum_{n}I_{n}n,
\label{In-first}\\
\frac{2\pi}{\phi_{0}}
(\langle n^{4}\rangle-3\langle n^{2}\rangle)|_{\phi=0} & = &
\beta\sum_{n}I_{n}n^{3}.
\label{In-second}
\end{eqnarray}
If we assume $I_n$ for $n>2$ are negligibly small,
we can solve Eq. (\ref{In-first})
and (\ref{In-second})
for $I_{1}$ and $I_{2}$,
\begin{eqnarray}
I_{1} & = & \frac{1}{3\beta}(-\langle n^{4}\rangle
+ 3\langle n^{2}\rangle^{2} + 4\langle n^{2}\rangle)|_{\phi=0},
\label{defofi1}\\
I_{2} & = & \frac{1}{6\beta}(-\langle n^{4}\rangle
+ 3\langle n^{2}\rangle^{2} + \langle n^{2}\rangle)|_{\phi=0}.
\label{defofi2}
\end{eqnarray}
Therefore we can calculate approximately the first and the second harmonics
of persistent current from expectation values
$\langle n^{2k}\rangle$ at $\phi=0$.
The charge stiffness is proportional to the mean-square winding numbers:
$ \rho_s=L\partial^{2}F/\partial\phi^{2}|_{\phi =0} =%
(L/\beta)\langle n^{2}\rangle|_{\phi =0}$.

Since our Lagrangian is invariant under the global transformation
$\theta \rightarrow\theta +2\pi k$ with $k$ being an integer,
the $\theta$'s could go far away from the initial values
during the updating processes of MC simulations.
In order to prevent this behavior which is inconvenient to
the simulations, we added a mass term $(m/2)\theta^2$
to the Lagrangian. After some experiments we found $m=0.02$ is
appropriate and we used this value in the following numerical
calculations.
We chose $g_{\tau}=1/\sqrt{2}$
which corresponds approximately to quarter filling
in the microscopic model of spinless fermions
with nearest neighbor interactions \cite{haldane1}.
We discretized the system in spatial and imaginary time
directions, so that the system size and the inverse temperature
are characterized by integers $L$ and $L_{\tau}$, respectively.
In the simulations we fixed the system size $L=8$ and
the particle density $\rho_0=0.25$.
We took 10,000 warm-up steps and 60,000 MC steps for measurements.
The impurity average was taken over 300-600 realizations
using the flat probability bounded by $W$: $-W<V_{\rm imp}(x)<W$.
Since there is a term $\langle n^2 \rangle^2$ in Eq. (\ref{defofi1})
and (\ref{defofi2})
we made two independent MC runs for each impurity realization
in order to minimize the statistical errors in the impurity average.

Figures \ref{fig1} show the first harmonics $I_{1}$ and charge stiffness
$\rho_{s}$ as functions of impurity potential strength $W$
for different interaction strengths. Temperature
is fixed to $1/L_{\tau}=1/24$.
We can see the crossover behavior between weak and strong impurity limits.
As known previously, persistent current and charge stiffness are
independent of interactions when there is no impurity ($W=0$)
\cite{bouzerar,muller,loss}.
When $W$ is small, both $I_{1}$ and $\rho_{s}$ are suppressed
as $g_{x}$ increases, {\it i.e.}, as repulsive interactions increase,
whereas they increase with $g_{x}$ in the strong impurity
region ($W>2$).
We mentioned earlier that the crossover should occur because the system
resides in the Luttinger liquid regime for weak impurities
whereas the particles are localized with strong impurities.
Therefore we believe the crossover occurs when the localization
length reaches the system size.

The temperature dependence of $I_1$ and $\rho_s$
are shown in Figs. 2 and 3. We can see that $I_1$ and $\rho_s$
have peaks at certain temperatures. This peak behavior comes from
two competing effects: depinning by thermal activation
and temperature destruction of quantum coherence. The former
effect is important at rather low temperature but the latter dominates
at higher temperature. This behavior has been seen in a single
impurity problem for a Wigner crystal \cite{krive}, and here
we find the essentially the same feature in the Luttinger
liquid in the presence of finite density of impurities.
The peak position $T^*$ is expected to be determined by those competing
effects, that is,
$T^*$ should be controlled by the effective impurity strength.
As discussed above, the effective impurity strength
in the weak impurity region is enhanced by interactions.
Therefore $T^*$ should be larger for larger $g_x$.
Actually this is what we can see in Figs. \ref{fig2} (a) and (b).
Figures. \ref{fig3} (a) and (b) show that, in strong impurity region,
$T^*$ is shifted to the lower side as $g_x$ increases,
because interactions would reduce the impurity strength.

Another interesting feature is the role of interactions
at rather high temperature.
For weak impurities, interactions always suppress the persistent
current and the stiffness in the whole temperature range.
When we have strong impurities, however, interactions
would screen the impurities only at low temperature, and
would suppress the current and the stiffness at high temperature.
This could be because electrons are already delocalized
by the thermal activation at high temperature
but the CDW correlation still remains, in other words,
the localization length becomes shorter than the system
size above some temperature although the CDW correlation
length is large. Therefore the role of
interactions in strong impurity region at high temperature
is expected to be the same as in weak impurity region.

In summary, we have reported on the first MC simulation of the
1D ring problem to see the interplay between disorder and
electron-electron interactions.
We have shown a crossover behavior from the weak-impurity
to strong-impurity region. When the impurities are weak, interactions
try to suppress the persistent current and charge stiffness.
On the other hand, interactions enhance the current and the stiffness
in the presence of strong impurities. In other words,
interactions increase the effective strength of weak impurities,
and decrease the effective strength of strong impurities.
We gave a physical
picture of these opposite behaviors, and we believe the crossover
occurs when the localization length becomes order of the system size.
The temperature dependences of persistent current and charge
stiffness are also calculated. We showed they have peaks
at certain temperatures and the peaks shift as we change
interaction strength.
The shifts are consistent with the
change of the effective impurity strength by interactions.
At high temperature the thermal activation would screen
the impurity effect and interactions act in the same way as
they do with weak impurities.

The author would like to acknowledge useful discussions with
S. M. Girvin, C. Hanna, K. Moon, H. Weber, and G. S. Canright.
The work was partially supported by NSF Grant No. DMR-9416906.

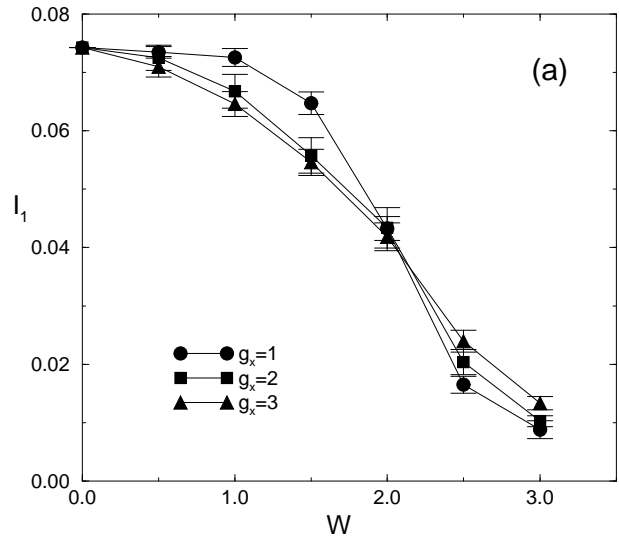
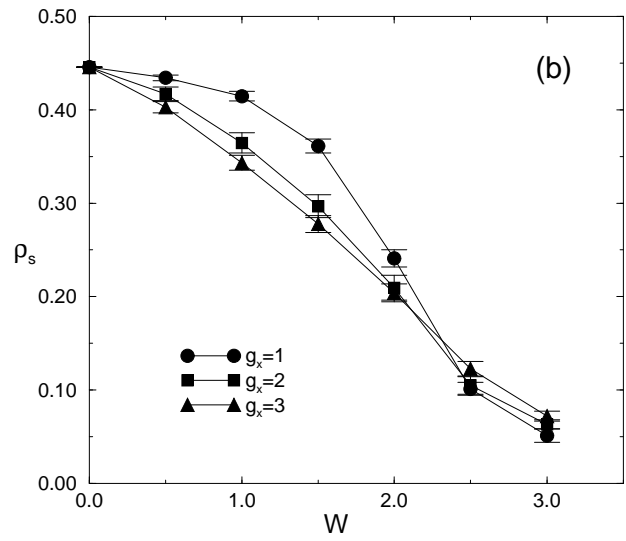
\begin{figure}
\caption{(a) Impurity and interaction dependence of the first
harmonics $I_1$ of persistent current. $g_{x}$ is a renormalized
parameter which is an increasing function of the microscopic
particle-particle interactions.
(b) The same dependence of charge stiffness $\rho_s$.}
\label{fig1}
\end{figure}
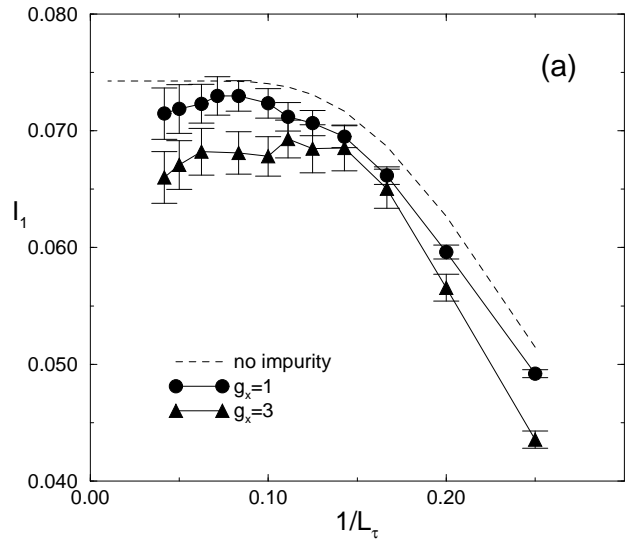
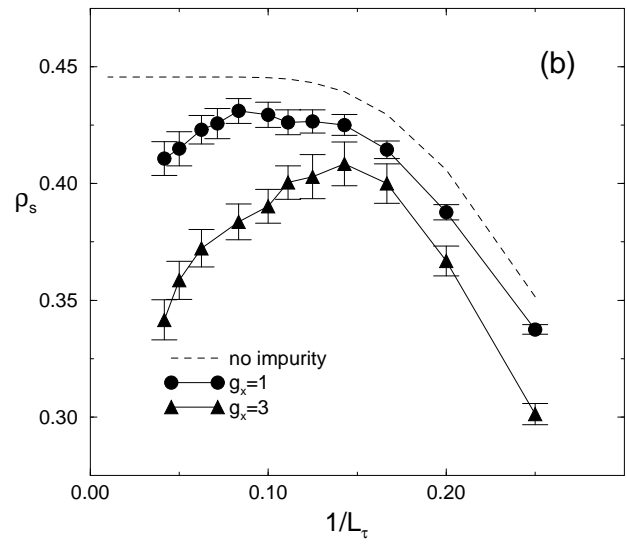
\begin{figure}
\caption{Temperature dependence of $I_1$ (a) and $\rho_s$ (b)
in the weak-impurity region (W=1). The dashed lines are for a
Luttinger liquid with no impurity.}
\label{fig2}
\end{figure}
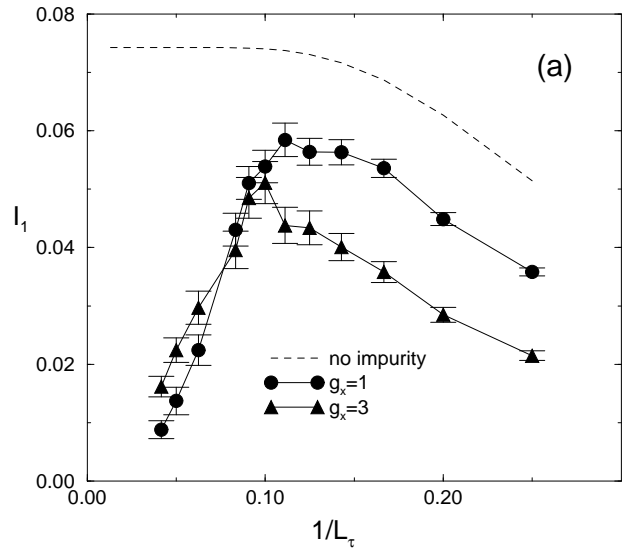
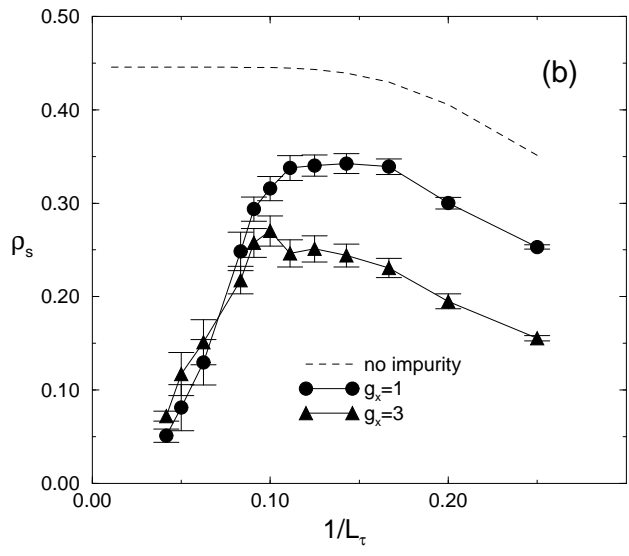
\begin{figure}
\caption{Temperature dependence of $I_1$ (a) and $\rho_s$ (b)
in the strong-impurity region (W=3).}
\label{fig3}
\end{figure}

\begin{references}
\bibitem{this}
This is true since our system is finite. In infinite chains, however,
electrons are localized no matter how weak impurities are,
at least when interactions are repulsive (See Ref. \cite{giamarchi}).
\end{references}
\end{document}